\documentclass{aa}  
\input{psfig.sty}  
\newcommand{\beq}{\begin{equation}}  
\newcommand{\eeq}{\end{equation}}  
\newcommand{\lapprox}{$\stackrel {<}{_{\sim}}$}  
\def\alp{\mbox{$\alpha$}}  
\def\farcd{\hbox{$.\mkern-4mu^\circ$}}

\def\arcmin{\hbox{$^\prime$}}  
\def\arcsec{\hbox{$^{\prime\prime}$}}  
\def\solar{\mbox{$_{\normalsize\odot}$}}  
  
\def\deg{\hbox{$^\circ$}}  
  
\begin{document}  
\title{OB Stellar Associations in the Large Magellanic  
Cloud:\\ Survey of young stellar systems}  
  
\author{D. Gouliermis\inst{1}  
\and M. Kontizas\inst{2}  
\and E. Kontizas\inst{3}  
\and R. Korakitis\inst{4}}  
  
\offprints{Dimitrios Gouliermis\\  
\email{dgoulier@astro.uni-bonn.de/horus@mailbox.gr}}  
  
\institute{Sternwarte der Universit\"{a}t Bonn, Auf dem H\"{u}gel  
        71, D-53121 Bonn, Germany  
        \and Department of Astrophysics Astronomy \& Mechanics, Faculty  
of Physics, University of Athens, GR-15783 Athens, Greece  
        \and Institute for Astronomy and Astrophysics, National  
Observatory of Athens, P.O. Box 20048, GR-118 10 Athens, Greece  
        \and Dionysos Satellite Observatory, National Technical  
University of Athens, GR-15780 Athens, Greece  
}  
  
\date{Received 20 January 2003 / Accepted 25 March 2003}  
  
\abstract{  
The method developed by Gouliermis et al. (2000, Paper I), for  
the detection and classification of stellar systems in the LMC, was used  
for the identification of stellar associations and open clusters in the  
central area of the LMC. This method was applied on the stellar catalog  
produced from a scanned 1.2m UK Schmidt Telescope Plate in U with a field  
of view almost 6\farcd5 $\times$ 6\farcd5, centered on the Bar of this  
galaxy. The survey of the identified systems is presented here followed by    
the results of the investigation on their spatial distribution and their  
structural parameters, as were estimated according to our proposed  
methodology in Paper I. The detected open clusters and stellar  
associations show to form large filamentary structures, which are often  
connected with the loci of HI shells. The derived mean size of the stellar  
associations in this survey was found to agree with the average size found  
previously by other authors, for stellar associations in different 
galaxies. This common size of about 80 pc might represent a universal 
scale for the star formation process, whereas the parameter correlations 
of the detected loose systems support the distinction between open 
clusters and stellar associations.  
}  
  
\titlerunning{Survey of Young Stellar Systems in the LMC}  
\authorrunning{D. Gouliermis et al.}  
\maketitle  
  
\section{Introduction}  
  
The spatial distribution and the statistical study of the properties of  
young stellar systems in a galaxy provide useful information on its  
structure and its formation history. Almost all the stars are born as  
members of various kinds of stellar groups, stellar associations being a 
very interesting one (Gomez et al. 1993; Massey et al. 1995). The variety  
of sizes of such stellar systems suggests a hierarchy in the formation of  
stellar structures as was earlier noted by McKibben Nail \& Shapley  
(1953). In addition, the investigation of a large sample of star forming 
regions in nearby galaxies allows us to study the behaviour of star 
formation in galactic scale.  Such a study has been done with use of the 
spatial investigation of the star formation history throughout the whole 
body of dwarf galaxies (e.g. Dohm-Palmer et al. 1997) and the distribution 
of the most active centres of recent star formation in distant galaxies 
(e.g. Hunter et al. 1998).  
  
The Large Magellanic Cloud (LMC) shows to be an ideal laboratory for the  
study of different stellar populations, due to its wide variety of stellar  
systems, it is relatively close to us (Madore \& Freedman 1998), there is  
low dust extinction (Harris et al. 1997) and we observe it almost face on  
(Westerlund 1997), while its depth seems to be very small (Caldwell \&  
Couslon 1986). In the LMC the star formation mechanisms in galactic scale  
have been investigated through the spatial distribution of Cepheids in  
this galaxy by Elmegreen \& Efremov (1996). This investigation was focused  
on a time scale of recent star formation \lapprox\ $5 \times 10^{8}$ yr.  
Harris \& Zaritsky (1999), expanding the research of stellar formation to  
larger areas and toward earlier times, give a quantitative description  
of the distribution of different stellar populations in the LMC.  They  
found indications of hierarchy in the formation of young stellar groups  
for length-scales between about 30 and 550 pc. As far as the initial  
spatial distribution of newly born stellar populations and their evolution  
concerns, it seems that these populations are mostly confined in small  
scale structures as was recently found in the SMC by Maragoudaki et al.  
(2001).  
  
In order to investigate the spatial distribution of the clustered young
populations in a galaxy one must identify the most recent formed stellar
systems. Stellar associations being typically loose stellar systems
(Blaauw 1964)  characterised by their bright blue populations, are
considered tracers of the distribution of the youngest population in a
galaxy in a specific galactic scale of about 80 pc (Efremov \& Elmegreen
1998). Still, their detection is not a trivial task. According to Hodge
(1986) the differences of the samples of stellar associations in various
galaxies arise from the use of different observational material and
selection criteria. He concludes that {\it with only the integrated colour
and patchiness of the images as criteria the danger of having severe
selection effects in the sample is acute} and he notes that the problem of
identifying stellar associations must be approached by carrying out proper
controls and statistical techniques in order to be sure that we are
dealing with physical groupings. Following these advises we developed an
objective method for the detection of loose young stellar systems in the
LMC (Gouliermis et al. 2000, from here on Paper I). The method is based on
the one proposed by Kontizas et al. (1994) and the selection criteria
presented by Kontizas et al. (1999).
  
In this paper we present the results of our investigation on the detection
of loose stellar systems, in the total area of about $6\farcd5 \times
6\farcd5$ around the Bar of the LMC. This area according to Westerlund
(1997) covers a subsystem, characterised by young stellar populations (see
also Nikolaev \& Weinberg 2000). Our investigation here consists of the
following steps: (1) Detection of stellar systems in the whole area of a
1.2m UK Schmidt Telescope plate in U by applying the method and the
criteria of Paper I (see Sect. 2). (2) Estimation of the structural
parameters of the detected systems according to the assumptions of Paper I
(Sect. 3.1). (3)  Classification of the systems according to their stellar
density into bound, intermediate and unbound systems (Sect. 3.2),
following the classification scheme proposed in Paper I. (4) In Sect. 3.3,
comparison of our survey with other catalogs and objects found in the same
areas. (5)  Study of the spatial distribution of the detected unbound and
intermediate systems in the LMC, which takes place in Sect. 4. (6)
Statistical study of the properties of these systems (Sect. 5). General
conclusions are given in Sect. 6.
  
\section{Detection of Stellar Systems}  
  
\subsection{Plate Description}  
  
We applied the detection method on the whole stellar catalog of a
photographic direct plate taken with the 1.2m UK Schmidt Telescope in U.  
The ID of the plate is U12346 and its centre coordinates are 05$^{\rm h}$
25$^{\rm m}$ 36$^{\rm s}$, $-$69{\deg} 35{\arcmin} 30{\arcsec} (in J2000)  
with a field of view of about 6\farcd5 $\times$ 6\farcd5. It was exposed
for 150 minutes on January 17, 1988. A UG1 filter was used on a
hypersensitized Kodak IIIa-J emulsion.  The stellar catalog was produced
by the Automated Plate Measuring Facility (APM) in Cambridge.  The data
were calibrated and checked for completeness (see Paper I).  This plate
covers the central area of the galaxy around the Bar, where most of the
associations found by Lucke \& Hodge (1970) (from here on LH) are located.  
In addition it covers a large area southern of the LH survey, where many
new systems were identified.
  
\begin{figure*}  
\centerline{\hbox{  
\psfig{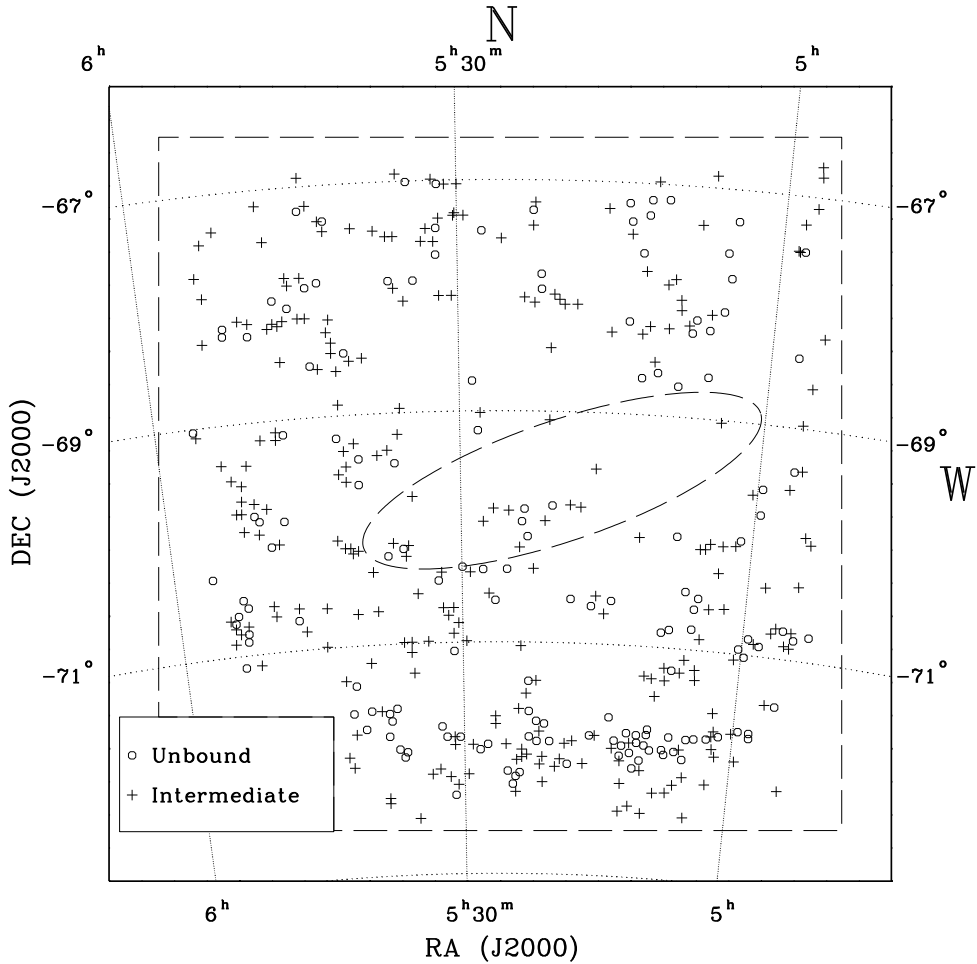}  
}}  
\caption{Survey of the intermediate and unbound systems detected in U in  
the area of 6\farcd5 $\times$ 6\farcd5 around the LMC Bar. The points  
represent the densest centres of the systems. The limits of the plate and  
the area of the Bar are marked with dashed lines.}  
\label{survey}  
\end{figure*}  
  
\subsection{Application of the Method}  
  
The method is based on star counts on a square grid, as we describe in  
Sect. 3 of Paper I.  We repeat the star counts six times, each for a  
different range of magnitudes (magnitude slice), in which the stellar  
catalog was divided and one time for all stars brighter than $\simeq$ 17  
mag (cumulative catalog). The adopted magnitude slices, as well as  
representative values of the mean background surface stellar density and  
its standard deviation $\sigma$ for different areas of the plate, are  
listed in Table 2 of Paper I. We applied the star counts procedure on 52  
different quadrilateral areas with dimensions varying from 0\farcd3 to  
2\farcd0, which cover the whole region of the plate.  The selection of the  
boundaries of these areas was made so that each one has a uniform  
distribution of field stars, meaning a more or less uniform background  
surface density. For each area we used the corresponding mean background  
density and its standard deviation ($\sigma$) in order to reveal the  
stellar concentrations which appear with a density exceeding the  
background by a specific limit in $\sigma$s (density threshold) on the  
cumulative catalog, satisfying the first criterion set in Paper I (Sect.  
3).  According to the second criterion each concentration should appear in  
at least two magnitude slices, one of them being for bright stars, at the  
same position as in the cumulative catalog.  
  
The method depends on two parameters. The first is the resolution limit,  
which corresponds to the grid element size for counting stars and sets the  
minimum size of the detected systems. The second is the lower density  
difference (in $\sigma$s) of each low-density grid element from its  
neighbours so that it can be accepted as ``member'' of the concentration,  
in order different neighbouring systems lying within the same density  
boundaries to be separated during the detection. More details are given in  
Sect. 3.1 of Paper I. The method's performance concerning these  
parameters, as well as for different density thresholds was thoroughly  
checked in Sect. 3.2 of Paper I. As a consequence the grid element size  
was selected to correspond to 20 pc $\times$ 20 pc. The lower density  
difference of neighbouring grid elements in order to be accepted as members  
of the same concentration was set to 1$\sigma$ and the lower density of  
the grid elements, which represent statistical important stellar groups  
(density threshold) was set to 3$\sigma$, both above the local background  
surface density.  
  
\section{The Detected Systems}  
  
\subsection{Parameters Estimation}  
  
Structural parameters for the detected systems were estimated following  
the calculations and assumptions of Sect. 4.1 of Paper I. The size of each  
system was defined as the maximum dimension of a quadrilateral enclosing  
its 3$\sigma$ density boundaries. The Spitzer radius $r_{\rm Sp}$ was  
computed as the mean square of the stars' distances from the point of peak  
density within this quadrilateral. The average central stellar density of  
each system (in M{\solar} pc$^{\rm -3}$) was estimated within the radius  
where half the number of its stars is contained, which may represent very  
well the half-mass radius of the system. The mass estimation of the  
systems needed a more thorough investigation and could only by achieved based  
on several assumptions.  The difficulties in such estimation and the  
assumptions used are discussed in detail in Paper I. In general, in order  
to estimate the total mass of each detected system we had to transform the  
luminosities of the stars from our catalog into masses by using isochrone  
models from Alongi et al. (1993) and to extrapolate the number of the  
brighter stars to the lower mass end due to incompleteness of the data. It  
was assumed that all detected systems are young (with ages around 5 Myr)  
and that their Mass Functions have slopes varying from $\Gamma \simeq  
-$1.0 to $-$1.6. Considering that the estimated masses depend on the  
errors of counting stars and mainly on the span of the adopted MF slopes  
and age, the extracted estimation errors can be as large as an order of a  
magnitude.  
  
\begin{table*}
\caption{Sample table of data and estimated parameters available for the
detected intermediate and unbound stellar systems. A part of the 
catalog of the unbound systems is shown here. The full catalogs are 
available from CDS.} 
\begin{center}
\begin{tabular}{cccrccccc}
\hline
%&&&&&&&& \\
ID&RA&DEC&Size&$N_{\star}$&$r_{\rm Sp}$&$r_{\rm h}$&Mass&$\rho$\\
&\multicolumn{2}{c}{(J2000)}&(pc)&&(pc)&(pc)&($\times10^{3}$ M{\solar})&
(M{\solar} pc$^{\rm -3}$) \\
\hline
\hline
%&&&&&&&& \\
%1&2&3&4&5&6&7&8&9 \\
%&&&&&&&& \\
%\hline
A015&05 22.0&$-$71 40&  83&  57 $\pm$  8& 26.1& 24.9& 1.8 - 3.5& 0.05 
$\pm$ 0.02\\
A016&05 20.5&$-$71 51&  83& 116 $\pm$ 11& 33.0& 30.7& 3.4 - 6.9& 0.05 
$\pm$ 0.02\\
A017&05 22.8&$-$71 48& 125& 166 $\pm$ 13& 60.2& 50.6& 4.9 - 9.8& 0.01 
$\pm$ 0.01\\
A018&05 22.9&$-$71 20&  63&  36 $\pm$  6& 22.8& 19.9& 1.1 - 2.3& 0.06 
$\pm$ 0.03\\
A019&05 21.1&$-$71 42&  83&  65 $\pm$  8& 30.0& 27.0& 2.0 - 4.0& 0.04 
$\pm$ 0.02\\
A020&05 16.0&$-$71 47&  42&  11 $\pm$  3& 12.1& 12.3& 0.4 - 0.8& 0.10 
$\pm$ 0.05\\
A021&05 18.4&$-$72 02&  42&  24 $\pm$  5& 21.0& 21.4& 0.8 - 1.6& 0.03 
$\pm$ 0.02\\
A022&05 23.8&$-$72 07&  63&  51 $\pm$  7& 37.7& 40.5& 1.6 - 3.2& 0.01 
$\pm$ 0.00\\
A023&05 22.8&$-$71 35&  63&  18 $\pm$  4& 20.9& 16.6& 0.6 - 1.2& 0.06 
$\pm$ 0.03\\
A024&05 13.9&$-$71 37&  63&  38 $\pm$  6& 28.9& 25.5& 1.2 - 2.4& 0.03 
$\pm$ 0.01\\
A025&05 21.9&$-$71 51&  63&  29 $\pm$  5& 27.3& 30.2& 0.9 - 1.9& 0.01 
$\pm$ 0.01\\
A026&05 01.6&$-$71 43& 145& 208 $\pm$ 14& 58.7& 53.6& 6.1 -12.2& 0.01 
$\pm$ 0.01\\
A027&05 04.3&$-$71 46& 187& 302 $\pm$ 17& 66.6& 61.4& 8.7 -17.5& 0.01 
$\pm$ 0.01\\
A028&05 02.9&$-$71 45&  62&  47 $\pm$  7& 25.9& 24.2& 1.5 - 2.9& 0.04 
$\pm$ 0.02\\
A029&05 06.4&$-$71 53& 187& 271 $\pm$ 16& 72.9& 68.1& 7.8 -15.7& 0.01 
$\pm$ 0.00\\
A030&05 05.2&$-$71 46& 166& 289 $\pm$ 17& 61.5& 53.9& 8.3 -16.7& 0.02 
$\pm$ 0.01\\
A031&05 09.6&$-$71 45& 125&  75 $\pm$  9& 53.9& 44.5& 2.3 - 4.6& 0.01 
$\pm$ 0.00\\
A032&05 12.6&$-$71 57& 145& 177 $\pm$ 13& 66.5& 62.4& 5.2 -10.4& 0.01 
$\pm$ 0.00\\
A033&05 10.8&$-$71 46&  83&  39 $\pm$  6& 36.0& 26.2& 1.2 - 2.5& 0.03 
$\pm$ 0.01\\
A034&05 09.8&$-$71 51& 125&  92 $\pm$ 10& 48.6& 35.2& 2.8 - 5.5& 0.02 
$\pm$ 0.01\\
A035&05 07.6&$-$71 55& 125& 119 $\pm$ 11& 67.8& 62.0& 3.5 - 7.1& 0.01 
$\pm$ 0.00\\
\hline
\end{tabular}
\end{center}
\label{partab}
\end{table*}

\subsection{Classification of the Detected Systems}  
  
We used the central stellar density of each system ($\rho$) in order to  
classify it as bound ($\rho \geq 1.0$ M{\solar} pc$^{\rm -3}$),  
intermediate ($0.1 < \rho < 1.0$ M{\solar} pc$^{\rm -3}$) or unbound  
system ($\rho \leq 0.1$ M{\solar} pc$^{\rm -3}$). This classification  
scheme is proposed in Sect. 4.1 of Paper I, where a discussion on the  
stability of stellar systems according to these density limits is also  
presented.  In total 494 stellar systems were found in the area of  
6\farcd5 $\times$ 6\farcd5 around the LMC Bar in the U plate. From them 82  
were classified as bound, 259 as intermediate and 153 as unbound systems.  
Our interest here is focused on the intermediate and unbound detected systems. 
They represent loose young stellar concentrations, which our method was  
designed for. Consequently almost all the detected bound systems were  
found with sizes on the detection limit of 20 pc and they are of no  
particular importance in this investigation. The detected intermediate and
unbound systems and their parameters are available in two tables from CDS
(one table for every category). Table \ref{partab} here is a sample of the
data given in these tables. In Col. 1 the identification number of each  
system is given. Cols. 2 and 3 show the coordinates of the dynamical  
centre of the system, which is taken to be the most dense part of the  
system. The size of each system is given in Col. 4, and in Cols. 5,6  
and 7 we present the number of bright stars $N_{\star}$ counted within the  
system's boundaries, its Spitzer radius $r_{\rm Sp}$, as well as its  
half-mass radius $r_{\rm h}$. The mass limits and the mean central stellar 
density of each system is given in Cols. 8 and 9 respectively. The  
identification numbers of the intermediate systems have the prefix `O',  
since the large majority of these systems most probably represents Open  
Clusters. For the unbound systems, considering that they are mostly  
Stellar Associations we used the prefix `A'. This numbering scheme will be  
used from here on.  
  
During the investigation on the detected systems in two selected areas of  
U plate, as was presented in Paper I, it was found that according to the  
spectral classification, all the identified unbound systems show to have  
the characteristics of OB stellar associations. Considering the various  
uncertainties in the estimation of the parameters of the systems, we also  
found that at least 30\% of the intermediate systems are probably stellar  
associations, since they show to have stellar densities very close to the  
limit of $\rho \simeq$ 0.1 M{\solar} pc$^{-3}$, while $\sim$ 68\% of them  
showed an excess of bright OB stars and we accepted them as candidate  
stellar associations or open clusters.  Taking these under consideration  
we present the results of our investigation on the detected intermediate  
and unbound systems in the whole U plate. The map of the survey of these  
systems is shown in Fig. \ref{survey}.    
  
\begin{table*}
\caption{Nomenclature, acronyms and references on the surveys of various
kinds of objects found to be connected to our survey according to
SIMBAD.}
\begin{center}
\begin{tabular}{llllc}
\hline
Object& SIMBAD &Acronyms&References&Number of\\   
Type& Nomenclature&&&objects\\
\hline
\hline
Globular Cluster&GlC&NGC      & Sulentic \& Tifft 1973&1\\
&&H88&Hodge 1988&2\\
\hline
Cluster of Stars&Cl*&SL&Shapley \& Lindsay 1963&105\\
&&NGC&Sulentic \& Tifft 1973&53\\
&&LT&Lortet \& Testor 1984&1\\
&&H88&Hodge 1988&19\\
&&[BH88]&Bhatia \& Hatzidimitriou 1988&8\\
&&BMG &Bhatia \& MacGillivray 1989&1\\
&&KMHK&Kontizas et al. 1990&87\\
&&BRHT&Bhatia et al. 1991&10\\
&&ZHT AN&Zaritsky et al. 1997&3\\
&&[BE99]&Battinelli \& Efremov 1999&4\\
&&BSDL&Bica et al. 1999&50\\
\hline
Open (galactic) Cluster&OpC&HS&Hodge \& Sexton 1966&20\\
&&NGC&Sulentic \& Tifft 1973&3\\
&&BMG &Bhatia \& MacGillivray 1989&17\\
&&BRHT&Bhatia et al. 1991&4\\
\hline
Association of Stars&As*&LH&Lucke \& Hodge 1970&60\\
&&NGC&Sulentic \& Tifft 1973&17\\
&&BMG &Bhatia \& MacGillivray 1989&1\\
&&KKDAB&Kontizas et al. 1994&1\\
&&BSDL&Bica et al. 1999&219\\
\hline
HII (ionized) region&HII&LHA 120-N&Henize 1956&28\\
&&PKS&Shimmins \& Day 1968&1\\
&&NGC&Sulentic \& Tifft 1973&2\\
&&DEM L&Davies et al. 1976&60\\
&&FHW95&Filipovic et al. 1995&1\\
\hline
Nebula of unknown nature&Neb&NGC&Sulentic \& Tifft 1973&2\\
&&BSDL&Bica et al. 1999&67\\
\hline
%HI (neutral) region&HI&[MNM83]&McGee et al. 1983&3\\
%\hline
%HI shell&sh&KDS99&Kim et al. 1999&2\\
%\hline
Super-giant shell&&SGsh-LMC&Meaburn 1980&2\\
\hline
Star Forming Region&&Shapley-&Shapley 1956&6\\
\hline
Molecular Cloud&MoC&LMC-CO&Morgan 1992&2\\
&&[CK96]&Caldwell \& Kutner 1996&2\\
&&[KRB97]&Kutner et al. 1997&2\\
&&[JGB98]&Johansson et al. 1998&4\\
\hline
\hline
\end{tabular}
\end{center}
\label{simbad}
\end{table*}

\subsection{Comparison with Other Surveys}  
  
In this section we present the results of the comparison of our survey
with other already published catalogs of stellar systems in the LMC. This
comparison is made in order to test the efficiency of our method in
comparison with others, concerning the number of the detected systems,
their positions and their estimated sizes.  We state in the introduction,
that the results of the detection of a specific type of stellar systems is
directly connected to the method and data used.  Thus, the detection
criteria adopted are very important, if we are to compare the results of
two different detection methods. The most complete available catalog of
stellar systems in the LMC is presented by Bica et al. (1999). This survey
covers the whole area of the galaxy and contains almost 6500 clusters,
associations and nebulae. Still we are not going to compare our results
with those of Bica et al., due to our differences with these authors
concerning the definition of a stellar association and in consequence the
adopted criteria. We found though all the Bica et al.  (1999) objects
related to our systems as is shown later.
  
The method of Bica et al. (1999) as was presented by Bica \& Schmitt  
(1995) is based on identification by eye on ESO/SERC R and J Sky Survey  
Schmidt films. In addition the distinction between clusters and  
associations according to these authors ``is based primarily on the  
stellar density, but additional criteria are the magnitude distribution of  
stars and the occurrence of irregular shape, which characterise  
associations'' (Bica \& Schmitt 1995). It seems that the detection  
criteria used was purely qualitative.  Consequently systems classified by  
these authors as associations do not meet systematically the criteria,  
which characterise this type of systems (see e.g. Kontizas et al. 1999).  
On the other hand the definition of stellar associations (roughly as loose  
concentrations of bright blue stars) used by Lucke \& Hodge (1970) is in  
line with the typical criteria for stellar associations and the criteria  
of our method. Thus we compare our results here with the LH catalog of LMC  
associations due to the similarities of the two methods, as far as the  
criteria concerns.  
  
In the area covered by the U plate there are 102 LH associations (Lucke \&  
Hodge 1970). We checked the coincidence of our identified systems with the  
LH associations for each smaller area, on which we applied the detection  
method. For each area we constructed the cumulative isodensity contour map  
(like the maps of Fig. 4 in Paper I), so that the limits of each detected  
system to be drawn after the limits of the LH associations of the same  
area were overplotted. The comparison was carried out by eye on these  
maps. We applied this kind of comparison due to the fact that the systems  
in both surveys have not symmetrical appearance, so it is possible that a  
centre to centre comparison wouldn't give any useful results. We accepted  
that two systems coincide even when these systems overlap each other with  
no centre coincidence. There are also cases where one system in the one  
catalog overlaps more than one systems of the other.  
  
\begin{table*}
\caption[]{Sample of the catalog of objects found by SIMBAD 
located in areas covered by the intermediate and unbound systems 
of this survey. The full catalog is available upon request
to the authors.}
\begin{center}
\begin{tabular}{cllccl}
\hline  
%&&&&& \\
ID&Simbad &Additional Related&RA&DEC&Type\\
  &Object &Objects&\multicolumn{2}{c}{(J2000)}&\\
%&&&&& \\
\hline  
\hline
%&&&&& \\
%1&2&3&4&5&6\\
%&&&&& \\
%\hline
 A015&BSDL 1416 &              &05 21 59&$-$71 41.3&As*  \\                        
     &BSDL 1424 &in BSDL 1428  &05 22 05&$-$71 41.9&As*  \\                        
     &BSDL 1428 &              &05 22 08&$-$71 40.8&As*  \\                        
 A016&BSDL 1304 &              &05 20 09&$-$71 53.2&As*  \\                        
     &BSDL 1321 &              &05 20 32&$-$71 53.4&As*  \\                        
 A017&KMHK 846  &              &05 23 02&$-$71 45.4&Cl*  \\                        
     &BSDL 1419 &              &05 21 59&$-$71 48.3&As*  \\                        
     &BSDL 1475 &              &05 22 41&$-$71 48.1&As*  \\                        
     &BSDL 1532 &              &05 23 30&$-$71 48.2&Cl*  \\                        
 A018&BSDL 1483 &in DEM L 164  &05 22 50&$-$71 21.1&As*  \\                        
 A019&KMHK 824  &              &05 21 31&$-$71 43.1&Cl*  \\                        
     &BSDL 1354 &              &05 21 09&$-$71 41.9&As*  \\                        
     &BSDL 1372 &              &05 21 19&$-$71 43.9&As*  \\                        
     &BSDL 1369 &              &05 21 20&$-$71 41.9&As*  \\                        
 A020&HS 202    &= KMHK 709    &05 16 03&$-$71 48.3&OpC  \\                        
     &BSDL 1088 &LHA 120-N 194 in it &05 15 57&$-$71 47.8&As*  \\                        
 A023&DEM L 165 &= LHA 120-N 198&05 22 27&$-$71 35.9&HII  \\                        
     &BSDL 1479 &in LHA 120-N 198 &05 22 46&$-$71 36.2&As*  \\                        
     &BSDL 1493 &in LHA 120-N 198 &05 22 56&$-$71 36.0&Cl*  \\                        
 A025&KMHK 829  &              &05 21 46&$-$71 51.1&Cl*  \\                        
 A026&BSDL 460  &              &05 01 48&$-$71 43.2&As*  \\                        
 A027&NGC 1840  &in BSDL 586   &05 05 18&$-$71 45.7&Cl*  \\                        
     &BSDL 586  &              &05 04 52&$-$71 46.2&As*  \\                        
     &BSDL 604  &in BSDL 586   &05 05 23&$-$71 47.1&As*  \\                        
 A029&KMHK 551  &= SL 235      &05 06 23&$-$71 49.1&Cl*  \\                        
 A030&NGC 1840  &= KMHK 529    &05 05 18&$-$71 45.7&Cl*  \\                        
     &BSDL 586  &NGC1840 in it &05 04 52&$-$71 46.2&As*  \\                        
     &BSDL 604  &in BSDL 586   &05 05 23&$-$71 47.1&As*  \\                        
 A034&BSDL 757  &              &05 09 10&$-$71 52.4&As*  \\                        
\hline
\end{tabular}
\end{center}
\label{smdbsmpl}
\end{table*}
  
We identified 73 LH associations. The 29 unidentified LH associations are
located in crowded regions characterised by high nebulosity, which makes
their identification a rather difficult task. Among them are the easily
recognized but highly nebulous LH 58 (Garmany et al. 1994) and the LH
associations located near 30 Doradus (Parker 1993; Parker \& Garmany
1993). More specifically, five of the LH associations not detected by us
are located within and around the area of 30 Doradus (LH 81, LH 90, LH
100, LH 105 \& LH 113), while additional seven are located in the dense
area northwest of and outside the Bar (LH 58, LH 67, LH 71, LH 73, LH 74
\& LH 85). In addition thirteen of the undetected LH associations belong
to the group of the 19 LH stellar associations, which cover the central
area and the northwest edge of the Bar. From these we detected LH 11, LH
12, LH 20, LH 39, LH 44 \& LH 59, which are all located very close to the
limits of the Bar. Finally our method did not detect four small LH
associations: LH 5, LH 82, LH 88 and LH 122. We conclude, thus, that the
method did not perform well in crowded regions and in regions where
nebulosity is very high. This result is not due to the detection method
itself, but due to the nature of the data used. It is worth noting that in
the U filter dense nebula regions like the 30 Doradus and the Bar are
shown to be extremely bright. Consequently, the detection of stars by the
scanning machines is affected by this phenomenon and the resulting stellar
catalogs in these regions are incomplete.  Indeed in the areas covering
the Bar, where we performed our method, we noticed phenomena of bad
statistics. The U plate on the other hand is one of the most appropriate
for the detection of young stellar populations, which is the aim of this
survey. In addition the coincidence of our detected systems with LH
associations in the rest areas of the plate was excellent.

In general we were able to detect 412 intermediate and unbound systems in
the same area where Lucke \& Hodge detected 102 associations, meaning that
this survey is almost four times richer than the one by Lucke \& Hodge.
Although the latter relied upon various criteria in addition to star
density, the central criterion of both methods is the appearance of
stellar associations as loose concentrations of bright young stars. So,
one may ask why this difference between the two surveys. Considering that
stellar density plays important role in both methods it seems that the
higher number of systems in our survey is probably due to the ability of
our method to detect loose stellar concentrations in areas where the eye
is not able to identify any significant stellar group unless there is
nebulosity related.

We confirmed the results of the comparison of the positions of the systems
in our survey with those of the LH associations using SIMBAD\footnote{\sc
SIMBAD: Set of Identifications, Measurements and Bibliography for
Astronomical Data} ({\tt http://simbad.u-strasbg.fr/}). We searched for
all the known stellar systems (as well as for other objects, as HII
regions, nebulae, etc.), which are located around the centre of each one
of our systems within a search radius equal to the half of the size of the
system as it was estimated in Sect. 3.1.  It should be taken under
consideration that this kind of ``centre to centre'' search, within
specific search radii, gave excellent results as far as the coincidence of
unbound systems and LH associations concerns, verifying the results of our
search. On the other hand, the search in SIMBAD for coincidence of known
objects with the detected intermediate systems verified only two thirds of
the coincidence of intermediate systems with LH associations found by us,
probably due to the systematically smaller search radii, which were used
in SIMBAD for these systems.  As it is said earlier, there are cases,
where more than one of our systems coincide with a LH association.
Probably these LH associations are stellar aggregates, for which we
detected the various internal subsystems. There are also cases, where more
than one LH associations were found to coincide with one of the systems of
our survey. Such a case is the one of LH 93, LH 94, LH 97 \& LH 98 west of
30 Doradus, which Lucke \& Hodge (1970) mark within the limits of one
larger association (LH 96), and which we detected through two systems
(A110 and O168), one of them being a large stellar association.
  
\begin{figure}  
\centerline{\hbox{  
\psfig{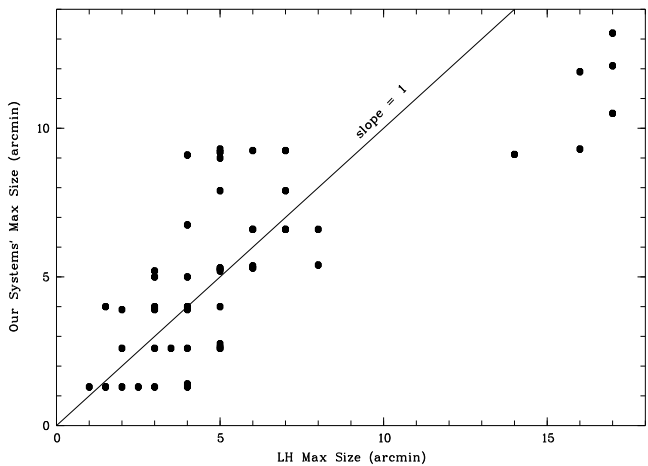}  
}}  
\caption{Comparison of the maximum sizes of the systems found in both our  
and LH surveys.}  
\label{melhsz}  
\end{figure}  
  
In Fig. \ref{melhsz} we compare the maximum sizes of our systems with the
ones of the corresponding LH associations. In cases where more than one
systems of one catalog coincide with only one of the other, we compared
the size of this system with the total size of all the subsystems
coinciding with it. As is shown from Fig. \ref{melhsz} the coincidence is
very good, considering a gap for sizes between $\sim$ 10 and 15 arcmin.  
This gap is due to the lack of systems with sizes in this range in the LH
catalog only.  The two groups of systems shown in Fig.  \ref{melhsz} give
a linear correlation with slopes 0.9$_{-0.2}^{+0.1}$ for the small systems
($<$ 10 arcmin) and 0.95$_{-0.5}^{+.05}$ for the larger ones ($>$ 10
arcmin). These values give a very good coincidence of our systems' sizes
with the ones of LH associations, being very close to the value of
excellent coincidence (slope=1) also shown in Fig. \ref{melhsz}.  In the
case of LH associations with sizes less that 10 arcmin the correlation
show to follow this line, considering the relatively wide spread in sizes.  
In the case of the larger systems, though the slope is nearly unit, the
zero point of the fitted line is well above zero, moved by about 5 arcmin.
This implies that the `larger' systems as was detected by us are
systematically smaller that the corresponding LH associations by about 5
arcmin. Probably this is due to two factors: (1) The sizes as were
estimated here are based purely on star counts and we did not consider the
related nebulosity, as Lucke \& Hodge did. (2) The correspondence of the
systems is not always one to one, since as we already stated there are
cases where there are more than one systems of our catalog corresponding
to one system in the LH catalog and vise versa.  So, the comparison of the
overall size of all these systems to the size of the corresponding one can
not be precise.
  
\begin{figure*}[t]  
\centerline{\hbox{  
\psfig{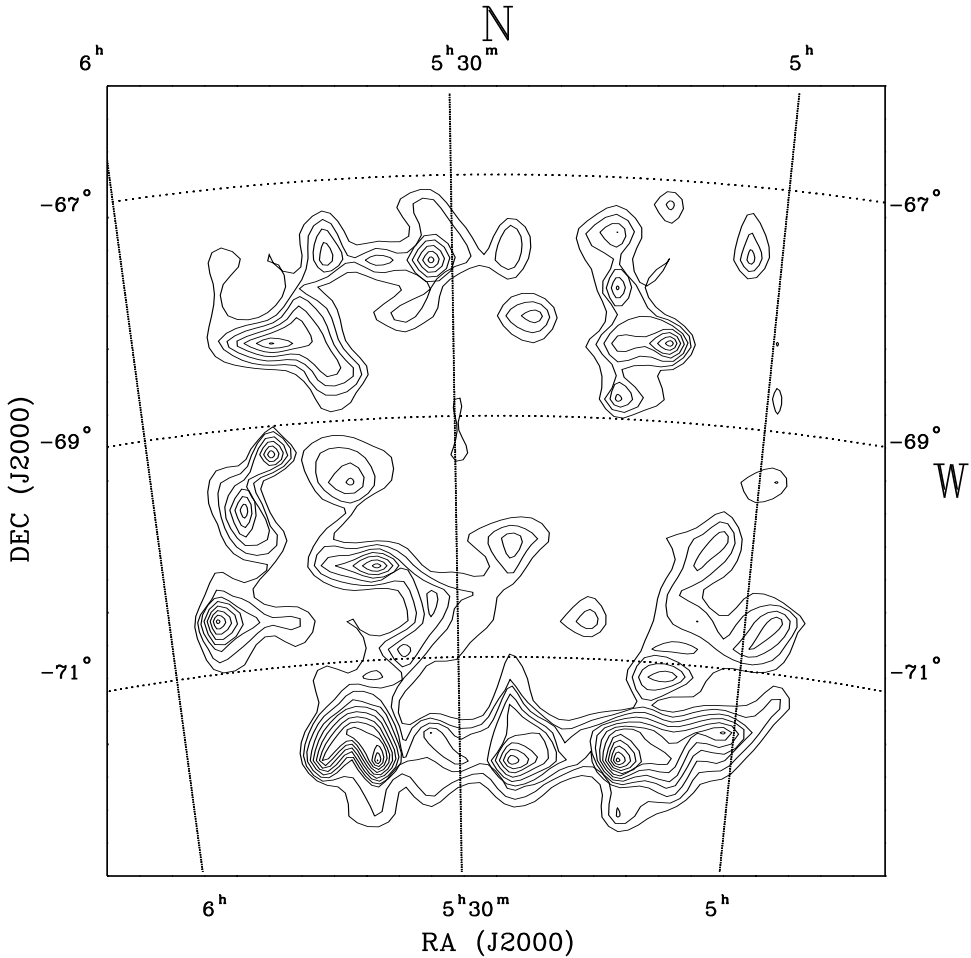}  
}}  
\caption{Isodensity contour map produced by counting both the intermediate  
\& unbound systems. These systems show  filamentary structures  
(with some compact parts) in the external regions around the  
Bar of the LMC.}  
\label{cmap23}  
\end{figure*}  
  
As a second step to our search for objects related to our systems through  
SIMBAD we used the catalogs by Lucke \& Hodge (1970) and Bica et al.  
(1999) to find additional ``inter-relations'' between these  
objects, as well as to find new objects. In Table \ref{simbad} we present  
summarised results of our SIMBAD search. We found 9 different types of  
non-stellar objects given in Col. 1 of the table. The nomenclature and  
acronyms of these types are given in Cols. 2 and 3 respectively. In  
Cols. 4 and 5 we give the corresponding bibliographical references and  
the number of objects found per acronym. Maybe the most interesting  
results are the ones concerning clusters, associations and HII regions, as  
they were classified by other authors: 341 clusters, 298 associations and  
92 HII regions found in SIMBAD to coincide in location with the  
intermediate and unbound systems detected by us. Ofcourse one cannot  
exclude the chance of superposition of SIMBAD clusters on our  
associations. This can be proved from the Galactic open and globular  
clusters also found in SIMBAD. Still, the estimation of an accurate  
fraction for the cases of super-positioning is almost impossible. Anyhow,  
the coincidence of the loci of the systems of our survey with clusters and  
small associations found by other authors can be approached as a physical  
phenomenon, since these small systems may represent core clusters in our  
associations. In total 225 intermediate and unbound systems of our survey  
were found to be related to 818 non-galactic SIMBAD objects. We compiled a  
catalog of all these objects. This catalog contains the systems of our  
survey (Col. 1), the related SIMBAD objects and the relations between  
them (Cols. 2 and 3), their coordinates (Cols. 4 and 5) and the  
corresponding SIMBAD nomenclature (Col. 6). Table \ref{smdbsmpl} shows a  
small portion of this catalog, which is not presented in full here due to  
its size and which is available upon request to the authors.  
  
\subsection{Completeness of the Survey in the Bar}  
  
In order to quantify the incompleteness of our survey in the area of the  
LMC Bar, we use the LH survey. Since we know the total number of LH  
associations and of our systems in the whole region, as well as in the  
area of the Bar, we can extrapolate the number of the systems we detected  
in the whole plate to the expected number of detected systems in the Bar  
area. In the whole area of the plate except the Bar (according to its  
limits as are plotted in Fig. \ref{survey}) we detected 147 unbound and  
246 intermediate systems. In the same area there are 85 LH associations.  
If we limit our catalog in the area, which is covered by the LH survey,  
then we find 105 unbound and 195 intermediate systems. This suggests that  
our survey covers $\sim$ 1.2 times more unbound and $\sim$ 2.5 times more  
intermediate systems than the survey of LH associations. If we consider  
that the distribution of the systems in the area of the Bar is uniform and  
that the detection performs consistently in every area of the plate, since  
the area of the Bar includes 17 LH associations, then we should expect to  
detect $\sim$ 21 unbound and $\sim$ 39 intermediate systems. We detected 6  
unbound and 14 intermediate systems, which gives a completeness of 35\%  
for the unbound and of 27\% for the intermediate systems in the area of  
the Bar.  
  
\section{Spatial Distribution of the Systems}  
  
From the map of Fig. \ref{survey} one may suspect the appearance of large
structures, which are formed by the systems. The incomplete detection of
systems in the area of the Bar produces the hole, which is shown in this
figure. The separate maps of the detected systems for each category show
these structures as well. We counted the detected intermediate and unbound
systems in quantrilateral grid with grid elements of sizes around 230 pc,
which is the size of a typical large stellar aggregate in the LMC (e.g.
Maragoudaki et al. 1998) and we produced the corresponding isodensity
contour map shown in Fig. \ref{cmap23}. For the production of the map we
didn't use any specific density threshold, since we used `clean' numbers
of systems not contaminated by any `background' contribution.
  
\begin{figure*}[t]
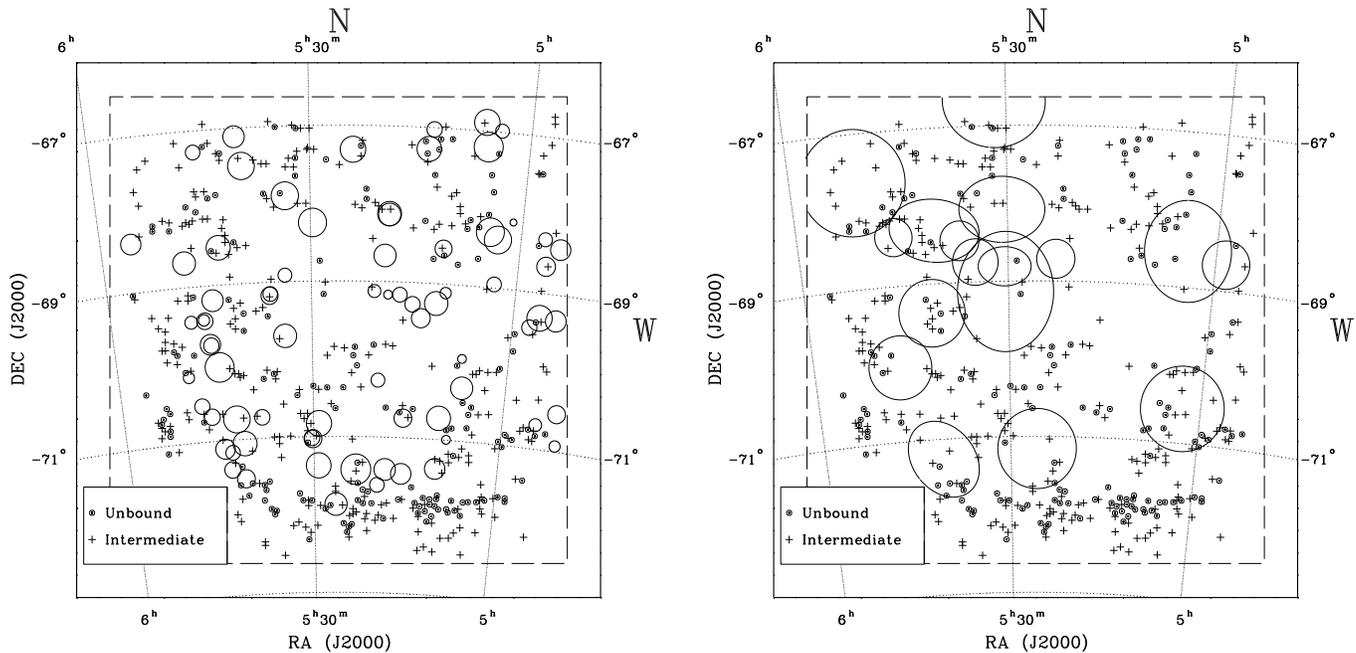
  
\centerline{\hbox{  
\psfig{figure=h4255f4a.ps,height=8.75truecm,width=9.75truecm,angle=270}  
\hspace*{-0.7cm}  
\psfig{figure=h4255f4b.ps,height=8.75truecm,width=9.75truecm,angle=270}    
}}  
\caption{The loci of the intermediate and unbound systems  on those of giant 
(left panel) and super-giant HI shells (right panel), as found by Kim et al.  
1999.}  
\label{wekim}  
\end{figure*}  
  
From the isodensity contour map of Fig. \ref{cmap23}, the existence of  
filamentary structures, which are formed by the detected systems is  
apparent. These structures show also to prefer the outer parts in the  
whole region. This could be due to the incomplete detection in the area of 
the Bar, but could also be due to the formation of the systems itself. If 
we consider that in the area of the Bar we detected only the 27\% of the  
expected intermediate systems, then one should expect to see a strong  
concentration of these systems in the area of the Bar. Still, this  
concentration would leave unaffected the structures shown on the NE and  
SW part of the map of Fig. \ref{cmap23}, due to the lack of systems in the  
areas between the Bar and these structures. This shows to be the case also  
for the unbound systems, of which we detected only the 35\% in the area  
of the Bar. Very interesting in the NNE part of the isodensity contour map  
is the arc structure, which is located right below of the super-bubble  
Shapley III (LMC 4), which several authors (e.g. de Boer et al. 1998;  
Efremov et al. 1998) have investigated concerning its star  
formation history. One may ask if other similar structures, which we  
observe in the isodensity contour map are connected to such star forming  
shells.  
  
More information on such an investigation can be given from observations  
on atomic hydrogen (HI) in the LMC, like the ones presented by Kim et al.  
(1999). These authors note that the structure of HI in the LMC is  
characterised by a large number of shells, as well as of filamentary and  
spiral structures. Taking under consideration that we detected stellar  
systems using data obtained in U, as well as the criteria chosen for the  
detection and acceptance of the systems as candidate stellar associations,  
then one should expect to find a relative coincidence of the spatial  
distribution of the systems in our catalog with the spatial distribution  
of atomic hydrogen, which is a good indicator of recent star formation.  
  
Indeed such a coincidence was found, when we overplotted the map of the
detected intermediate and unbound systems on the corresponding part of the
HI map of LMC by Kim et al. (1999). It was rather interesting that the
large filamentary structure in the SW part of our survey (see Fig.
\ref{survey} or Fig. \ref{cmap23}) found an almost perfect correspondence
in the survey of Kim et al. (1999). Another interesting result of this
comparison is that many of our systems are distributed around areas empty
in HI (holes). Kim et al. using their observations and additional
observations in H{\alp}, classified 103 candidate giant- and 23
supergiant-shells. We show in Fig. \ref{wekim} the comparison of the
positions of the stellar systems in our survey with the edges of the giant
(left panel) and supergiant (right panel) shells detected by Kim et al.  
This figure is an additional indication that probably some of the systems
in our survey are related to the boundaries of such shells. Consequently
the distribution of the systems forms arc-like structures shown in the
contour map of Fig. \ref{cmap23}. Could these systems be the result of
star formation events happening on the edges of HI shells? -Probably.

\begin{figure*}[t]
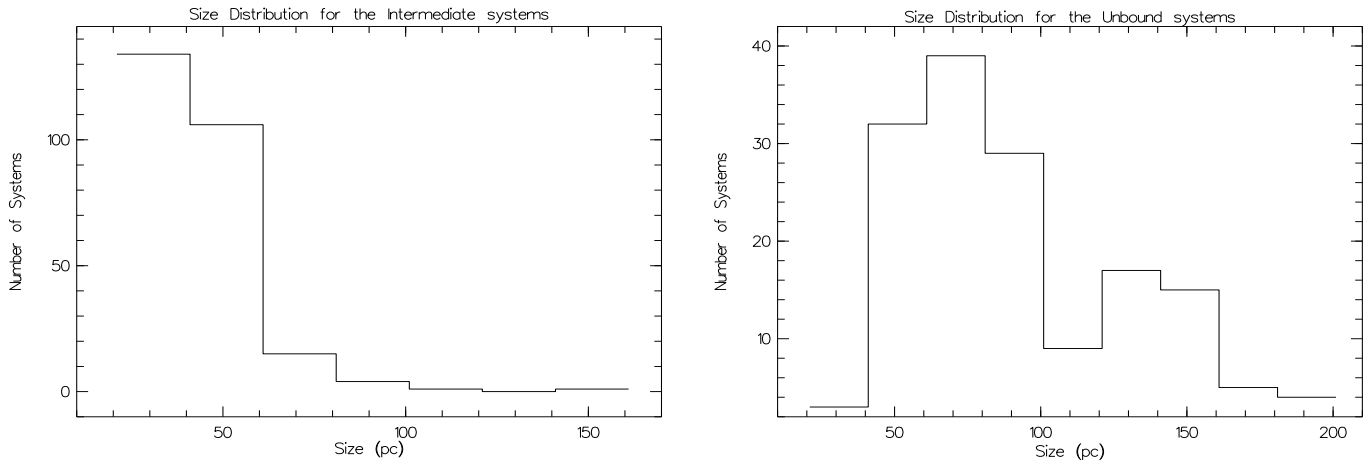
  
\centerline{\hbox{  
\psfig{figure=h4255f5a.ps,height=6.15truecm,width=9.75truecm,angle=270}  
\hspace*{-0.65cm}  
\psfig{figure=h4255f5b.ps,height=6.15truecm,width=9.75truecm,angle=270}  
}}  
\caption{Size distribution for the detected intermediate (left panel)  
and unbound systems (right panel).}  
\label{szdstr}  
\end{figure*}  
  
For example this seems to be the case for the systems found to coincide  
with the ``conjunction'' point of the supergiant shells in the NE part of  
the map (Fig. \ref{wekim} - right panel). Stellar systems of our survey  
were also found projected on the inner areas of some shells. Kim et al.  
comment that the shells which cover HII regions and known OB associations  
seem to expand faster. This fact implies an additional input of mechanical  
energy from active star forming regions.  Indeed we compared the areas of  
the HII regions found by SIMBAD to be related to systems of our survey and  
we found that more than half of them are located at the inner part of HI  
shells. It is interesting to note that the ages of the shells, as were  
estimated by Kim et al. (1999) are varying from about 2 to 11.5 Myr. This  
age estimation is in good agreement with the age of 5-10 Myr we assumed  
for the transformation of luminosities to masses for the detected systems  
during the estimation of their structural parameters.  
    
\section{The Systems' Parameters}  
  
\subsection{Sizes and Masses of the Systems}  
  
The sizes of the detected systems was measured as shown in Sect. 3.1. If  
we take all three categories of detected systems under consideration, we  
see that as one moves from the category of the most compact systems  
(bound) toward the less compact ones (unbound), the corresponding  
dimensions are getting larger with the bound systems having sizes close to  
our detection limit and the unbound representing the larger detected  
systems. More specifically as far as the intermediate systems concerns most  
of them have sizes clustered around 30 pc, while there are some systems  
with sizes up to 160 pc (Fig. \ref{szdstr} - left panel). The size  
distribution of the unbound systems shows a peak at about 70 pc and a  
second shorter peak at about 130 pc. Considering that the unbound systems  
found in Paper I are true stellar associations (as the spectral  
classification showed), then possibly all the unbound stellar systems of  
this survey represent the true population of stellar associations in the  
central area of LMC. The mean size of these systems was found around 85  
pc.  This result is in good agreement with the size distribution for  
stellar associations of galaxies in the Local Group, as was found by  
various investigators as will be shown below.  
  
In Table \ref{sasz} a compilation of existing data on detection methods  
and measured sizes is given for stellar associations in different  
galaxies. In Cols. 1 and 2 the name and the Hubble type of the galaxies  
are given. In Col. 3 we give the number of detected associations, while  
Col. 4 shows the minimum, average and maximum sizes of the detected  
associations. The corresponding references and methods are given in  
Cols. 5 and 6 respectively. This table shows the differences between  
galaxies as far as the number of detected associations and their sizes  
concerns. Hodge (1986), using a similar compilation of data available at  
that time, argues that the differences of the detected number of  
associations in various galaxies is a result of their greater distances,  
which makes the recognition of small associations more difficult and which  
leads to preferentially interpreting complexes of associations as single  
objects. Consequently, the associations of more distant galaxies are found  
to be larger.  
  
In addition, according to the same author the differences in the  
properties of stellar associations in different galaxies arise also from  
the use of different observational material and detection criteria. So the  
different sizes of associations in different samples could be a selection  
effect. Efremov (1988) found that the mean size of the OB associations  
($\simeq$ 80 pc) is a universal scale for the process of star formation in  
the galactic discs and Ivanov (1996) discuss the use of the  
average size of OB associations as a distance indicator of the parent  
galaxy. The same author found that the average diameter of the stellar  
associations of eight galaxies detected using CCD observations is 84 $\pm$  
15 pc, very close to Efremov's value. The average size of our detected  
unbound systems (86 pc) seems to be in a very good agreement with both  
these values. From the data of Table \ref{sasz} one can see that more than  
half of the surveys include associations with average sizes around the  
value of 80 pc (between 65 and 93 pc). Almost 30\% of the surveys have  
associations with sizes larger than about 114 pc up to 440 pc and there is  
only a small fraction (3 surveys) with smaller limits for the sizes of  
stellar associations (18 to 46 pc).  
  
\begin{table*}
\caption{Sizes of stellar associations in galaxies}
\begin{center}
\begin{tabular}{lcrrrrlc}
\hline
Galaxy& Hubble & Number &\multicolumn{3}{c}{Size (pc)}&References & Detection \\
Name  & Type   &        &min&mean&max&           & Method       \\
\hline
\hline
Sextans A& E   &   3 &   & 93&    & Ivanov 1996 &6 \\
M 31     & Sb  & 210 & 20& 80&    & Efremov et al. 1987 &1\\
         &     &  15 &   & 83&    & Ivanov 1996 &6 \\
NGC 7331 & Sb  & 142 &   &440&    & Hodge 1986  &1 \\
M 33     & Sc  & 143 &   &200&    & Humphreys \& Sandage 1980&1\\
         &     & 460 & 30& 80& 270& Ivanov 1987 &1 \\
         &     & 289 &  6& 66& 305& Ivanov 1991 &5 \\
         &     &   8 &   & 87&    & Ivanov 1996 &6 \\
         &     &  41 & 10& 40& 120& Wilson 1991 &3 \\
NGC 2403 & Sc  &  88 &160&348& 600& Hodge 1985a &1\\
NGC 4303 & SBbc& 235 &   &290&    & Hodge 1986  &1\\
LMC      & Irr & 122 &15 & 78& 150& Lucke \& Hodge 1970&1\\
         &     &2883 &  5& 18& 272& Bica et al. 1999&1\\
         &     & 153 & 21& 86& 190& This Paper      &7\\
SMC      & Irr &  70 &18 & 77& 180& Hodge 1985b      &1\\
         &     &  31 &50 & 90&270& Battinelli 1991 &4\\
         &     &$\sim$200&9&46&234& Bica \& Schmitt 1995 &1\\
NGC 6822 & Ir+ &  16 & 48&163& 305& Hodge 1977  &2\\
         &     &   6 &   & 72&    & Ivanov 1996 &6\\
IC 1613  & Irr &  20 & 68&164& 485& Hodge 1978  &2\\
         &     &   6 &   & 83&    & Ivanov 1996 &6 \\
Pegasus  & Irr &   3 &   & 65&    & Ivanov 1996 &6 \\
GR 8     & Irr &   3 &   &114&    & Ivanov 1996 &6 \\
Ho IX    & Im  &   3 &   & 72&    & Ivanov 1996 &6 \\
\hline
\hline
\multicolumn{8}{c}{DETECTION METHODS EXPLANATIONS}\\
\hline
\multicolumn{8}{l}{1: Detection by eye on photographic plates or films}\\
\multicolumn{8}{l}{2: Detection by eye using star counts from 
photoelectric and photographic observations}\\
\multicolumn{8}{l}{3: `friends of friends' grouping algorithm on stars 
from CCD observations}\\
\multicolumn{8}{l}{4: `Path Linkage Criterion' applied on O-B2 stars 
selected from objective-prism observations}\\
\multicolumn{8}{l}{5: `cluster analysis' technique on stars from 
photographic observations}\\
\multicolumn{8}{l}{6: Automated `cluster analysis' technique on OB stars 
selected from CCD observations}\\
\multicolumn{8}{l}{7: Objective statistical method based on star counts from 
photographic stellar catalogs}\\
\hline
%\multicolumn{8}{l}{\footnotesize Hubble types from:
%``The Messier Catalog''
%(http://www.seds.org/messier/xtra/supp/gal-ttab.html)}\\
%\multicolumn{8}{l}{\footnotesize \& ``LEVEL5'' 
%(http://nedwww.ipac.caltech.edu/level5/Glossary/Essay\_hodge.html)}\\
\end{tabular}
\end{center}
\label{sasz}
\end{table*}

If we select from Table \ref{sasz} the surveys of stellar associations,  
which have been compiled with the use of more advanced detection  
techniques (not by eye) then we limit the sample of surveys of  
associations to half. Almost all of them are in agreement on the average  
size of stellar associations. Specifically all 12 except 2 surveys give  
the size of a stellar association to be between 65 and 93 pc; this gives  
an average size of stellar associations equal to 80 $\pm$ 3 pc, in  
excellent agreement with the universal scale of star formation as given by  
Efremov (1988). Thus, one could almost safely conclude that as far as the  
size of stellar associations concerns, this might represent a specific  
global length-scale of star formation in a galaxy. Still, there are  
differences between different surveys of associations (like in the total  
number of associations in a galaxy), which might as well depend on the  
total luminosity or Hubble type of the host galaxy (Hodge 1986).  
  
Concerning the mass estimation of each detected stellar system, it should
be noted that the MF slope of the systems was the most important
assumption and that the uncertainties of the estimated masses up to an
order of a magnitude are due to the span of the adopted MF slopes (see
Sect. 4.1, Paper I).  This range ($-1.6$ \lapprox\ $\Gamma$\ \lapprox\
$-1.0$) is in excellent agreement with the established IMF slopes of LH
associations by several detailed studies of their stellar content (LH
117/118:  Massey et al. 1989; 30 Dor area: Parker \& Garmany 1993; LH 58:
Garmany et al. 1994; LH 47/48: Oey \& Massey 1995; Will et al. 1997; LH
1/2/5/8: Parker et al. 2001). These slopes were found to be comparable to
each other and clustered around $\Gamma\ \simeq -1.5$.

From the results of the investigations above, we were able to test the
mass estimation for four of our detected systems: A129/130 (LH 47/48) and
O123/A092 (LH 117/118). The MF slopes of the coinciding LH associations
(LH 47/48: Oey \& Massey 1995; Will et al. 1997; LH 117/118:
Massey et al. 1989) were used for an estimation of their masses in the
range 0.8 \lapprox\ $M$/M{\solar} \lapprox\ 15, which was selected during
our detection method for the mass estimation of the identified systems.
The masses of the systems were found to be consistent with those of the
related LH associations as estimated by their MF slopes, giving us
confidence in the mass estimation of the detected systems of this survey.

\begin{figure}  
\centerline{\hbox{  
\psfig{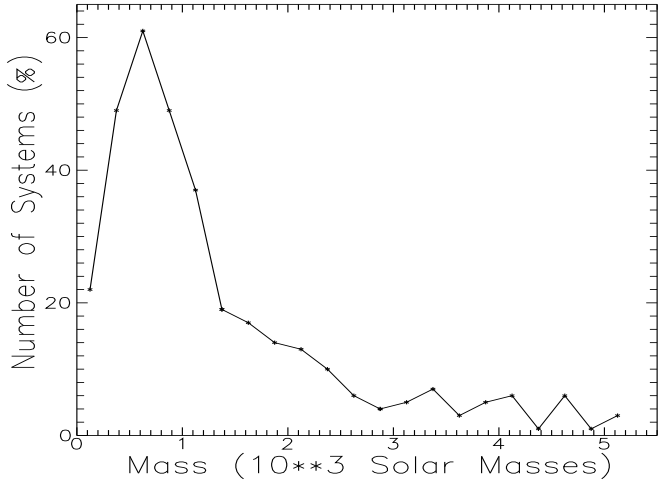}  
}}  
\caption{Frequency distribution of the masses of the detected intermediate 
and unbound systems. Masses are given in 10$^{3}$ M{\solar}, whereas the 
number of the systems are in percentage of the total detected systems.}  
\label{masdst}  
\end{figure}  
  
The frequency distribution of the masses found for the intermediate and
unbound systems is shown in Fig. \ref{masdst}. It seems to be a normally
distributed mass function around a mass of about 10$^{3}$ M{\solar}.  In
this figure the x-axis represents the average mass of the systems
estimated according to the assumptions of Sect. 4.1 of paper I and y-axis
the fraction of systems over the total of intermediate and unbound
detected systems (412 in total). We limit this graph at 5 $\times$
10$^{3}$ M{\solar}, which represents the majority (82\%) of the detected
systems. The remaining 18\% of the systems is divided in two groups: The
first 9\% includes systems with masses of 5 to 10 $\times$ 10$^3$
M{\solar} and the other with masses 10 to 64 $\times$ 10$^3$ M{\solar}.
The distributions of both groups are flat. It is interesting to point out
that this distribution implies that the mass function of the stellar
associations and open clusters in the LMC seems to be Gaussian.
  
\subsection{Parameter Correlations}  
  
\begin{figure*}[t]
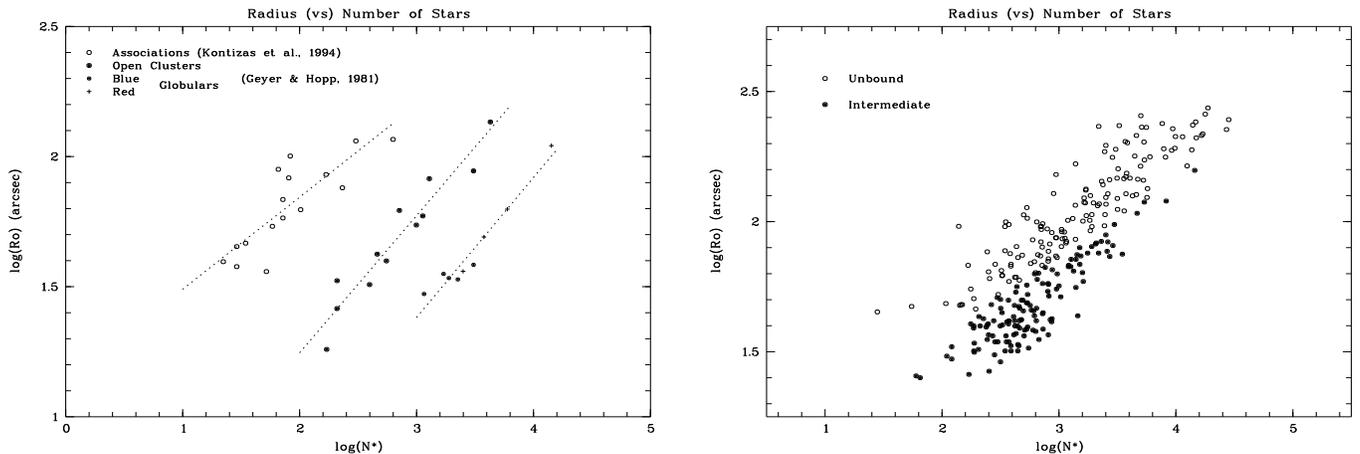
  
\centerline{\hbox{  
\psfig{figure=h4255f7a.ps,height=6.15truecm,width=9.75truecm,angle=270}  
\hspace*{-0.65cm}  
\psfig{figure=h4255f7b.ps,height=6.15truecm,width=9.75truecm,angle=270}  
}}  
\caption{Correlations of half-mass radii with the corresponding number of  
stars for stellar associations, open and globular clusters in the LMC  
(left panel) and for the intermediate and unbound systems of this survey  
(right panel).}  
\label{rh2nst}  
\end{figure*}

Geyer \& Hopp (1981) presented a correlation of the number of stars and  
the corresponding system's radius for 12 open and 8 globular clusters in  
the LMC. They found that the corresponding correlation coefficients are  
almost the same for both kind of systems. Still, the correlations for  
these two types of systems are parallel probably due to the systematic  
differences in number of stars per radius between them. In Fig.  
\ref{rh2nst} (left panel) the original diagram of Geyer \& Hopp (1981) is 
shown with the corresponding points for the stellar associations found by  
Kontizas et al.  (1994) overplotted. The correlation $r_{\rm  
h}$($N_{\star}$) for these systems is also almost parallel to the  
correlations of open and globular clusters. In Fig. \ref{rh2nst} (right  
panel) we also present the corresponding correlation of the  
half-mass radius to the included number of stars for the intermediate and  
unbound systems of this survey.  
  
In this diagram we observe that there is a differentiation in the radii of  
the systems at a limit of $r_{\rm h}$ $\sim$ 23 pc, meaning that there are  
no intermediate systems with radii larger than this limit, in contrast to  
the unbound systems. This limit (diameter of $\sim$ 46 pc) seems to meet  
the diameter limit as was defined by Efremov (1982) for clusters and  
stellar associations. The correlation of number of stars versus radius, as  
shown in the right panel of Fig. \ref{rh2nst} for both classes of systems  
follows almost an identical trend.  Specifically the relation $\log{r_{\rm  
h}}$ (vs)  $\log{N_{\star}}$ for the intermediate systems has a slope of  
0.35$\pm$0.02, while this slope for the unbound systems is 0.31$\pm$0.01.  
Both correlations can be fitted very well to a line. The statistical tests  
we performed for the goodness of fit, using Spearman and Pearson  
correlation coefficients, showed that indeed there is a strong correlation  
between the radius of a system and the corresponding number of stars. The  
correlation coefficients where found to be between 0.8 and 0.9.  
  
It is worth noting that these slopes are in a very good agreement with the  
one for the stellar associations found by Kontizas et al. (1994), which is  
equal to $\simeq$ 0.35$\pm$0.06. On the contrary the slopes for the open  
and globular clusters of Geyer \& Hopp (1981) are steeper equal to  
0.53$\pm$0.05 and 0.54$\pm$0.05 correspondingly. We performed additional  
statistical tests using the Kolmogorov-Smirnov method, in order to detect  
any possible similarities between the correlations for our systems and the  
ones of the Kontizas et al. associations and of the Geyer \& Hopp  
clusters. The only significant similarity found was the one between our  
intermediate systems and of the open clusters with 96\% of coincidence.  
  
It should be taken under consideration that the parameters used in these  
correlations are directly measured without any assumption to be used for  
their estimation.  So the number of stars used is the counted stellar  
number within each system's boundaries from the APM catalog, while $r_{\rm  
h}$ is the radius, where half of these stars are contained. Consequently  
these correlations are physically meaningful, as far as their slopes  
concerns. Still, they are to be taken arbitrarily due to the systematic  
incompleteness in the estimation of the actual numbers.  Interesting is also  
the fact that the distributions of associations, open clusters and  
globulars in the diagram of Geyer \& Hopp (1981) and Kontizas et al.  
(1994) systems are very well distinct to each other and this is probably  
due to incomplete samples.  The distributions of our larger sample, on the  
other hand, show that there is no clear distinction between systems of  
different types, which implies that there might be also ``hybrid'' systems  
between what we call associations and open clusters as far as these  
parameters correlations concerns.  
  
\subsection{Fainter Magnitude of Appearance}  
  
\begin{figure}  
\centerline{\hbox{  
\psfig{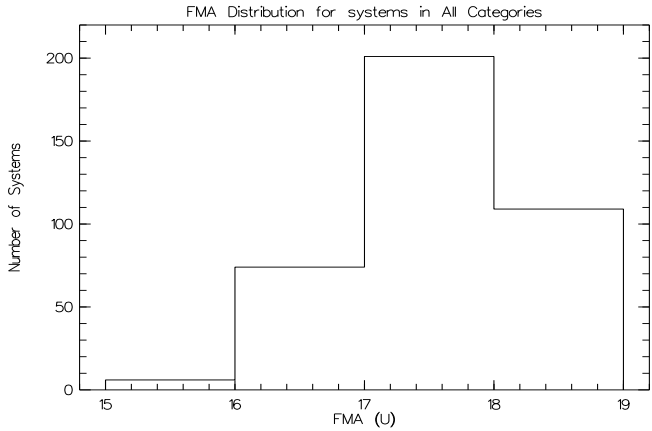}  
}}  
\caption{Number distribution of detected systems in all three categories  
according to their {\it ``Fainter Magnitude of Appearance''}.}  
\label{fmadst}  
\end{figure}

During the detection and study of stellar systems in two selected areas of  
the U plate, in Paper I, we observed that the fainter magnitude, where  
each systems first appears in the star counts differs from the one system  
to the other. We called this limit ``Fainter Magnitude of Appearance''  
(FMA). These differences in the FMA are apparent for systems of the same  
category, as well as for neighbouring systems. This phenomenon is now  
verified from the larger sample of systems in this paper. For this  
investigation we will also use the detected bound systems, since we are  
only interested in the fainter detected magnitude for every concentration  
of stars found, without taking any system category under consideration.  
The number distribution of all the detected systems according to their FMA  
is given in Fig. \ref{fmadst}. In this distribution we use steps of one  
magnitude in order to achieve statistically significant results, due to  
the relatively low number of systems. Considering that the detection limit  
of the plate is at around U $\simeq$ 20 mag, the incompleteness around the  
19th magnitude can be important. In this case one should expect a  
larger number of systems in the fainter magnitude bin. Still, the  
distribution of systems in different FMAs remains.  
  
Interesting seems to be also the spatial distribution of the detected  
systems according to their FMA. We found that as far as the FMA concerns  
the systems show to be concentrated in larger structures, most of which  
have filamentary shape and contains other smaller and denser  
concentrations of systems. As we note in Paper I this differentiation of  
the FMA should not depend of the density of the systems, since systems of  
different types show to have the same FMA. Probably it is also not due to  
intergalactic absorption, since the LMC reddening is not large enough to  
produce a magnitude differentiation up to four magnitudes (Harris et al.  
1997). In Paper I we give as a possible explanation for this phenomenon  
the existence of pre-main sequence stars with masses up to 8 M{\solar} in  
the systems, as was found by various authors for example in R 136 in 30  
Doradus (e.g. Sirianni et al. 1999). Still, this is a rough explanation  
and detailed analysis on selected young stellar systems in the LMC is  
needed in order to be verified.  
  
\section{Conclusions}  
  
In this paper we present the properties of the stellar associations and
open clusters in the LMC, as detected on a digitised, by APM, 1.2m UK
Schmidt Telescope Plate in U, using the method proposed in Paper I.  
There were found 494 stellar systems in an area of $\sim$ 6\farcd5
$\times$ 6\farcd5 around the LMC Bar.  We classified the detected systems
in three categories based on their stellar density in the half-mass
radius, as was estimated for every system according to the method and
assumptions of Paper I. There were found 82 bound systems with $\rho \geq$
1.0 M{\solar} pc$^{-3}$, 259 intermediate systems with 0.1 $\leq \rho
\leq$ 1.0 M{\solar} pc$^{-3}$ and 153 unbound systems with $\rho \leq$ 0.1
M{\solar} pc$^{-3}$.
  
Lucke \& Hodge (1970) detected in the same area 102 stellar associations.  
We detected almost the 70\% of them. The 29 non-detected LH stellar
associations are located in dense regions of ionised hydrogen and in the
Bar of the galaxy. In these regions the method was unable to detect low
concentration systems, due to the incompleteness of the stellar catalog.  
It was found that the coincidence of the sizes of the LH associations, as
were estimated by us with the ones of Lucke \& Hodge is very good. In
general we detected about four times more loose young stellar systems than
Lucke \& Hodge in their survey of associations. The higher number of
identified systems in our survey is probably due to the better detection
performance of our method in areas of loose stellar concentrations.  

Several studies have demonstrated the coevality of LH stellar associations
in their IMF slopes and star formation histories (e.g. Massey 1998; Massey
et al. 2000). The selection criteria of our method for the detection of
young loose stellar systems in the LMC were based on this coevality
concerning their MF slopes and stellar content. In addition, thorough
statistical tests were carried out for the best performance of the method
in detecting coeval loose concentrations of bright blue stars. These can
serve as evidence that the newly discovered unbound and intermediate
systems should be considered as candidates for true open clusters and
stellar associations and not just statistical coincidences.
  
The intermediate and unbound systems, as was found from their spatial
distribution, seem to construct large filamentary structures.  A
comparison of the loci of these systems with neutral hydrogen observations
shows that filamentary and arc structures are often formed from the
concentration of systems on the edges of giant and super-giant HI shells.  
The intermediate systems cover a size range of about 30 pc up to 160 pc.  
The size distribution of the unbound systems shows two peaks, the higher
first at 70 pc and the lower second at 130 pc.  The average size of these
stellar associations (86 pc) is in a very good agreement with the mean
size of stellar associations found in various galaxies (80 $\pm$ 3 pc),
which possibly represents a global length-scale of star formation.  The
mass function of the intermediate and unbound systems was found here to be
almost normally distributed around $10^{3}$ M{\solar}.
  
The correlation of the half-mass radius of every system to the  
corresponding number of stars, shows the existence of two different  
system types, the borders of which are overlapping each other. Still,  
there is a distinction between these two classes at a diameter limit of  
$\sim$ 46 pc. The differences in the distributions of $r_{\rm h}$ versus  
$N_{\star}$ for the intermediate and unbound systems of the survey  
indicates the different nature of these two kinds of systems, with the  
unbound representing dynamically loose stellar concentrations, which are  
expected to dissolve sooner than the intermediate, which seem to represent  
mostly open clusters.  
  
Throughout the whole catalog of detected systems the Fainter Magnitude of
Appearance (FMA) shows variations, which seem to be independent of the
system types and of the location of the systems. These variations are
found up to four magnitudes, which implies that the LMC absorption may not
be the reason and allowing us to suggest the existence of pre-main
sequence stars.

%\section*{Acknowledgements} 
\begin{acknowledgements} 
The authors would like to express their thanks to the 1.2m U.K. Schmidt
Telescope Unit for the loan of the plates and the APM \& SuperCOSMOS for
the digitisation of them. Our thanks are also due to the {\it General
Secretariat of Research and Technology} of the Greek Ministry of
Development and the {\it British Council} for their financial support, as
well as to the University of Athens (ELKE). This research has made use of
the SIMBAD database, operated at CDS, Strasbourg, France.
\end{acknowledgements}

\end{document}